\documentclass[preprint,showpacs,preprintnumbers,amsmath,amssymb]{revtex4}

\usepackage{graphicx}

\begin{document}

\title{
Post-Newtonian effects on Lagrange's equilateral triangular 
solution for the three-body problem 
}
\author{Takumi Ichita}
\author{Kei Yamada}
\author{Hideki Asada} 
\affiliation{
Faculty of Science and Technology, Hirosaki University,
Hirosaki 036-8561, Japan} 

\date{\today}

\begin{abstract}
Continuing work initiated in earlier publications 
[Yamada, Asada, Phys. Rev. D {\bf 82}, 104019 (2010), 
{\bf 83}, 024040 (2011)], 
we investigate the post-Newtonian effects on 
Lagrange's equilateral triangular solution 
for the three-body problem. 
For three finite masses, 
it is found that the equilateral triangular configuration 
satisfies the post-Newtonian equation of motion 
in general relativity, 
if and only if all three masses are equal. 
When a test mass is included, 
the equilateral configuration is possible 
for two cases: 
(1) one mass is finite and the other two are zero, 
or 
(2) two of the masses are finite and equal, and 
the third one is zero, namely a symmetric binary with a test mass. 
The angular velocity of the post-Newtonian equilateral triangular 
configuration is always smaller than the Newtonian one, 
provided that the masses and the side length are the same. 
\end{abstract}


\maketitle

\section{Introduction}
The three-body problem in the Newton gravity 
represents classical problems in astronomy and physics 
(e.g, \cite{Danby,Goldstein,Marchal}). 
In 1765, Euler found a collinear solution for 
the restricted three-body problem 
that assumes one of three bodies is a test mass. 
Soon after, his solution was extended for 
a general three-body problem by Lagrange, 
who also found an equilateral triangle solution 
in 1772. 
Now, the solutions for the restricted three-body problem 
are called Lagrange points $L_1, L_2, L_3, L_4$ and $L_5$, 
which are well known and described in textbooks 
of classical mechanics \cite{Goldstein}. 

Lagrange points have recently attracted renewed interests 
for relativistic astrophysics, 
where they have discussed the gravitational radiation 
reaction on $L_4$ and $L_5$ analytically \cite{Asada} 
and by numerical methods \cite{SM,Schnittman}. 

As a pioneering work, Nordtvedt pointed out that 
the location of the triangular points is very sensitive 
to the ratio of the gravitational mass to the inertial one 
\cite{Nordtvedt}. 
Along this course, it is interesting as a gravity experiment 
to discuss the three-body coupling terms at the post-Newtonian 
order, 
because some of the terms are proportional to a product of 
three masses as $M_1 \times M_2 \times M_3$. 
Such a triple product can appear only for relativistic three (or more) 
body systems but cannot for a relativistic compact binary 
nor a Newtonian three-body system. 

The relativistic perihelion advance of the Mercury 
is detected only after much larger shifts due to 
Newtonian perturbations by other planets such as 
the Venus and Jupiter are taken into account 
in the astrometric data analysis. 
In this sense, effects by the three-body coupling 
are worthy to investigate. 
Nevertheless, most of post-Newtonian works have focused on 
either compact binaries 
because of our interest in gravitational waves astronomy 
or 
N-body equation of motion (and coordinate systems) 
in the weak field such as the solar system (e.g. \cite{Brumberg}). 
Actually, future space astrometric missions 
such as Gaia 
\cite{GAIA,JASMINE}
require a general relativistic modeling of 
the solar system within the accuracy of a micro arc-second 
\cite{Klioner}. 
Furthermore, a binary plus a third body have been discussed 
also for perturbations of gravitational waves induced by the third body 
\cite{ICTN,Wardell,CDHL,GMH}. 

The theory of general relativity is currently 
the most successful gravitational theory describing 
the nature of space and time. 
Hence it is important to take account of general relativistic effects 
on three-body configurations. 
The figure-eight configuration that was found decades ago 
\cite{Moore,CM} has been numerically studied 
at the first post-Newtonian \cite{ICA} 
and also the second post-Newtonian orders \cite{LN}.
According to their numerical investigations, 
the solution remains true with a slight change 
in the figure-eight shape because of relativistic effects. 

On the other hand, the post-Newtonian collinear configuration 
has been recently obtained as a relativistic extension of 
{\it Euler's collinear one}, 
where three bodies move around the common center of 
mass with the same orbital period and always line up \cite{YA2010}. 
It may offer 
a useful toy model for relativistic three-body interactions, 
because it is tractable by hand without numerical simulations. 
The uniqueness of the collinear configuration 
has been also proven \cite{YA2011}. 

{\it Lagrange's equilateral triangular solution} has also 
a practical importance, since it is stable for some cases. 
Lagrange's points $L_4$ and $L_5$ 
for the Sun-Jupiter system are stable and indeed 
the Trojan asteroids are located there. 
Clearly it is of greater importance to investigate 
Lagrange's equilateral triangular solution 
in the framework of general relativity. 
Do the post-Newtonian effects admit such a triangular solution? 
No one doubts whether the particular configuration is still 
possible at the post-Newtonian order. 
We shall study this issue in this paper. 
The main purpose of this paper is to show that 
the equilateral triangular configuration 
can satisfy the post-Newtonian equation of motion, 
if and only if all three finite masses are equal.
Throughout this paper, we take the units of $G=c=1$.

\section{Newtonian Lagrange's equilateral triangular solution} 
First, we consider the Newton gravity among three masses 
denoted as $M_I$ $(I=1, 2, 3)$. 
The location of each mass is written as 
$\mbox{\boldmath $x$}_I$. 
We choose the origin of the coordinates, 
so that 
\begin{equation}
M_1 \mbox{\boldmath $x$}_1
+ M_2 \mbox{\boldmath $x$}_2 
+ M_3 \mbox{\boldmath $x$}_3 = 0 . 
\end{equation}
We start by seeing whether 
the Newtonian equation of motion for each body can be satisfied 
if the configuration is an equilateral triangle. 
Let us put $R_{12} = R_{23} = R_{31} \equiv a$, 
where we define the relative position between masses as 
\begin{equation}
\mbox{\boldmath $R$}_{IJ} 
\equiv \mbox{\boldmath $x$}_I - \mbox{\boldmath $x$}_J , 
\end{equation}
and $R_{IJ} \equiv |\mbox{\boldmath $R$}_{IJ}|$ 
for $I, J = 1, 2, 3$. 
Then, the equation of motion for each mass becomes 
\begin{equation}
\frac{d^2 \mbox{\boldmath $x$}_I}{dt^2} 
= - M \frac{\mbox{\boldmath $x$}_I}{a^3} , 
\label{N-EOM}
\end{equation}
where $M$ denotes the total mass $\sum_I M_I$. 
Therefore, it is possible that each body moves around 
the common center of mass with the same orbital period. 
Eq. (\ref{N-EOM}) gives 
\begin{equation}
\omega_N^2 = \frac{M}{a^3} , 
\label{omega2}
\end{equation}
where $\omega_N$ denotes the Newtonian angular velocity. 

\begin{figure}[t]
\includegraphics[width=12cm]{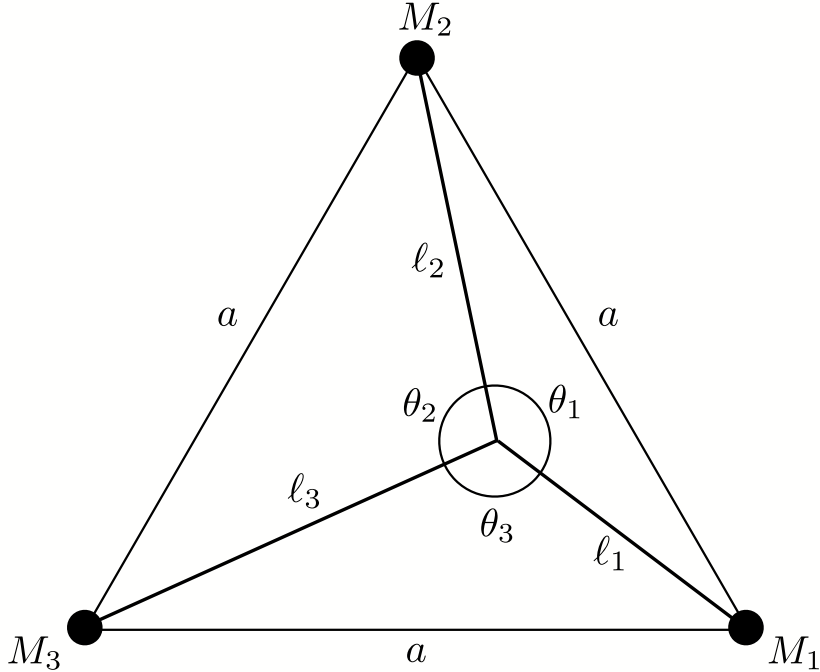}
\caption{ 
Equilateral triangular configuration. 
Each mass is located at one of the apices. 
We define $\theta_I$ ($I = 1, 2, 3$) 
with respect to the common center of mass. 
}
\label{f1}
\end{figure}

Figure \ref{f1} shows an equilateral triangular configuration. 
Let $\mbox{\boldmath $\ell$}_I$ 
denote the relative position vector of each mass 
with respect to the common center of mass 
(but not the geometrical center of the triangle) 
in the corotating frame with the angular velocity $\omega_N$. 
The angles between 
$(\mbox{\boldmath $\ell$}_2, \mbox{\boldmath $\ell$}_1)$,  
$(\mbox{\boldmath $\ell$}_3, \mbox{\boldmath $\ell$}_2)$ 
and 
$(\mbox{\boldmath $\ell$}_1, \mbox{\boldmath $\ell$}_3)$ 
are defined cyclically as 
$\theta_1$, $\theta_2$ and $\theta_3$ 
respectively. 
They are constant with time. 
There is an identity as $\theta_1 + \theta_2 + \theta_3 = 2\pi$, 
which can be used to delete one of the angles. 
The orbital radius $\ell_I \equiv |\mbox{\boldmath $\ell$}_I|$ 
of each body with respect to the common center of mass 
is obtained as \cite{Danby} 
\begin{eqnarray}
\ell_1&=& \frac{a}{M}\sqrt{M_2^2 + M_2M_3 + M_3^2} , 
\label{N-ell1}
\\
\ell_2&=& \frac{a}{M}\sqrt{M_1^2 + M_1M_3 + M_3^2} ,
\label{N-ell2}
\\
\ell_3&=& \frac{a}{M}\sqrt{M_1^2 + M_1M_2 + M_2^2} . 
\label{N-ell3}
\end{eqnarray}

\section{Post-Newtonian equilateral triangular solution}
Next, we consider the post-Newtonian effects on 
the triangular configuration 
by employing  
the Einstein-Infeld-Hoffman (EIH) equation of motion as \cite{MTW,LL,AFH} 
\begin{eqnarray}
\frac{d \mbox{\boldmath $v$}_K}{dt} 
&=& \sum_{A \neq K} \mbox{\boldmath $R$}_{AK} 
\frac{M_A}{R_{AK}^3} 
\left[
1 - 4 \sum_{B \neq K} \frac{M_B}{R_{BK}} 
- \sum_{C \neq A} \frac{M_C}{R_{CA}} 
\left( 1 - 
\frac{\mbox{\boldmath $R$}_{AK} \cdot \mbox{\boldmath $R$}_{CA}}
{2R_{CA}^2} \right) 
\right.
\nonumber\\
&&
\left. 
~~~~~~~~~~~~~~~~~~~~~
+ v_K^2 + 2v_A^2 - 4\mbox{\boldmath $v$}_A \cdot \mbox{\boldmath $v$}_K 
- \frac32 \left( 
\mbox{\boldmath $v$}_A \cdot \mbox{\boldmath $n$}_{AK} \right)^2 
\right]
\nonumber\\
&&
- \sum_{A \neq K} (\mbox{\boldmath $v$}_A - \mbox{\boldmath $v$}_K) 
\frac{M_A \mbox{\boldmath $n$}_{AK} \cdot 
(3 \mbox{\boldmath $v$}_A - 4 \mbox{\boldmath $v$}_K)}{R_{AK}^2} 
\nonumber\\
&&
+ \frac72 \sum_{A \neq K} \sum_{C \neq A} 
\mbox{\boldmath $R$}_{CA} 
\frac{M_A M_C}{R_{AK} R_{CA}^3} , 
\label{EIH-EOM}
\end{eqnarray}
where $\mbox{\boldmath $v$}_I$ denotes the velocity of each mass 
in an inertial frame 
and we define 
\begin{eqnarray}
\mbox{\boldmath $n$}_{IJ}&\equiv&
\frac{\mbox{\boldmath $R$}_{IJ}}{R_{IJ}} .
\end{eqnarray}

Let us see whether the three masses at the apices of 
an equilateral triangle 
can satisfy the EIH equation of motion. 
For such an equilateral triangle case, 
the second-order-mass terms are easy to handle, 
because every $R_{IJ}$ is the same as $a$. 
What we have to take care of is the velocity-dependent terms. 

We consider three masses in circular motion 
with the angular velocity $\omega$, 
so that each $\ell_I$ can be a constant. 
The position and velocity of each body are expressed as 
\begin{eqnarray}
\boldsymbol{x}_1 &=& \ell_1 
\begin{pmatrix}
\cos{\omega t} \\ \sin{\omega t}
\end{pmatrix} , 
\label{x1}
\\
\boldsymbol{v}_1 &=& \ell_1 \omega
\begin{pmatrix}
-\sin{\omega t} \\ \cos{\omega t}
\end{pmatrix} , 
\label{v1}
\\
\boldsymbol{x}_2 &=& \ell_2
\begin{pmatrix}
\cos{(\omega t + \theta_1)} \\ \sin{(\omega t + \theta_1)}
\end{pmatrix} , 
\label{x2}
\\
\boldsymbol{v}_2 &=& \ell_2 \omega
\begin{pmatrix}
-\sin{(\omega t + \theta_1)} \\ \cos{(\omega t + \theta_1)}
\end{pmatrix} ,
\label{v2}
\\
\boldsymbol{x}_3 &=& \ell_3
\begin{pmatrix}
\cos{(\omega t - \theta_3)} \\ \sin{(\omega t - \theta_3)}
\end{pmatrix} ,
\label{x3}
\\
\boldsymbol{v}_3 &=& \ell_3 \omega
\begin{pmatrix}
-\sin{(\omega t - \theta_3)} \\ \cos{(\omega t - \theta_3)}
\end{pmatrix} , 
\label{v3}
\end{eqnarray}  
where we used $\theta_1+\theta_2 = 2\pi - \theta_3$, 
$|\boldsymbol{x}_I| = \ell_I$ 
and $|\boldsymbol{v}_I| = \ell_I\omega$. 

For the later convenience, we compute the inner products between 
the velocity and relative position vectors as 
\begin{eqnarray}
\boldsymbol{R}_{12} \cdot \boldsymbol{v}_1
&=& - \ell_1 \ell_2 \omega \sin{\theta_1} ,
\label{inner1}
 \\
\boldsymbol{R}_{12} \cdot \boldsymbol{v}_2
&=& - \ell_1 \ell_2 \omega \sin{\theta_1} ,
 \\
\boldsymbol{R}_{12} \cdot \boldsymbol{v}_3
&=& \ell_3 \ell_1 \omega \sin{\theta_3}
- \ell_2 \ell_3 \omega \sin{(\theta_3 + \theta_1)} ,
 \\
\boldsymbol{R}_{23} \cdot \boldsymbol{v}_1
&=& \ell_1 \ell_2 \omega \sin{\theta_1}
- \ell_3 \ell_1 \omega \sin{(\theta_1 + \theta_2)} ,
 \\
\boldsymbol{R}_{23} \cdot \boldsymbol{v}_2
&=& - \ell_2 \ell_3 \omega \sin{\theta_2} ,
 \\
\boldsymbol{R}_{23} \cdot \boldsymbol{v}_3
&=& - \ell_2 \ell_3 \omega \sin{\theta_2} ,
\\
\boldsymbol{R}_{31} \cdot \boldsymbol{v}_1
&=& - \ell_3 \ell_1 \omega \sin{\theta_3} ,
 \\
\boldsymbol{R}_{31} \cdot \boldsymbol{v}_2
&=& \ell_2 \ell_3 \omega \sin{\theta_2} 
- \ell_1 \ell_2 \omega \sin{(\theta_2 + \theta_3)} ,
 \\
\boldsymbol{R}_{31} \cdot \boldsymbol{v}_3
&=& - \ell_3 \ell_1 \omega \sin{\theta_3} ,
 \\
\boldsymbol{v}_1 \cdot \boldsymbol{v}_2
&=& \ell_1 \ell_2 \omega^2 \cos{\theta_1} ,
 \\
\boldsymbol{v}_2 \cdot \boldsymbol{v}_3
&=& \ell_2 \ell_3 \omega^2 \cos{\theta_2} ,
 \\
\boldsymbol{v}_3 \cdot \boldsymbol{v}_1
&=& \ell_3 \ell_1 \omega^2 \cos{\theta_3} . 
\label{inner2}
\end{eqnarray}
Note that the inner product is defined in the 3-dimensional 
Euclid space 
as $\boldsymbol{\alpha} \cdot \boldsymbol{\beta} 
= \delta_{ij} \alpha^i \beta^j$ 
for $i, j = 1, 2, 3$. 
This is because the terms expressed by Eqs. (\ref{inner1})-(\ref{inner2}) 
appear at the post-Newtonian order but not 
at the Newtonian one.  
Therefore, it is sufficient to use the flat-space metric 
for computing these post-Newtonian terms. 
Corrections to the inner product in a curved spacetime 
make 2PN (or higher order) contributions and 
they are thus ignored in this paper. 

Let us explain how to describe an equilateral triangular
configuration in the Newtonian frame and the relativistic one. 
The same side length is given for the Newtonian triangle and 
the post-Newtonian one, so that each mass position 
can be denoted as the same vector $\boldsymbol{x}_I$ 
for both cases, whereas the position of mass center 
(not a geometric center) may be different. 
This treatment makes computations easier than 
another way that assumes the mass center position to be fixed as 
the same location for both frames by introducing  
three post-Newtonian position vectors $\boldsymbol{x}^{PN}_I$ 
(for the three masses) 
possibly different from 
the three Newtonian position vectors $\boldsymbol{x}^{N}_I$. 
Three position vectors are needed to calculate 
in the latter approach, whereas in this paper one position vector 
denoting the mass center is sufficient to specify the system.

In order to compute the orbital radius of each mass, 
the location of the mass center at the post-Newtonian order 
must be determined. 
It is expressed as \cite{MTW,LL} 
\begin{eqnarray}
\boldsymbol{G}_{PN} = E^{-1} \sum\limits_AM_A\boldsymbol{x}_A 
\left[1 + \frac12
\left(v_A^2 - \sum\limits_{B \neq A}\frac{M_B}{R_{AB}}
\right) \right] , 
\label{PN-COM}
\end{eqnarray}
where $E$ is defined as 
\begin{eqnarray}
E = \sum\limits_AM_A\left[1 + \frac12
\left(v_A^2 - \sum\limits_{B \neq A}\frac{M_B}{R_{AB}}
\right) \right] . 
\end{eqnarray}

By using Eq. (\ref{PN-COM}), 
the post-Newtonian orbital radius 
$\ell_{PN1} \equiv |\boldsymbol{x}_1 - \boldsymbol{G}_{PN}| 
\equiv \sqrt{\delta_{ij} (x_1^i - G_{PN}^i) (x_1^j - G_{PN}^j)}
$ 
is obtained as  
\begin{eqnarray}
\ell_{PN1}^2 &=& \ell_1^2 
 \nonumber \\
&& + \frac{a}{2M^3}  
( - 2M_1^2M_2^2 - 2M_3^2M_1^2 + 2M_1M_2^3 + 2M_3^3M_1 
+ M_2^3M_3 - 2M_2^2M_3^2
 \nonumber \\
&& 
~~~~~~~~~~~~~~
+ M_2M_3^3 - 2M_1^2M_2M_3 + M_1M_2^2M_3 + M_1M_2M_3^2 )  
\left(1 - \frac{a^3 \omega_N^2}{M} \right) . 
\label{PN-ell1}
\end{eqnarray}
By noting Eq. (\ref{omega2}), we find that 
the second term in the R.H.S. of Eq. (\ref{PN-ell1}) vanishes 
and hence $\ell_{PN1} = \ell_1$. 
By the cyclic permutations, we find also 
$\ell_{PN2} = \ell_2$ and $\ell_{PN3} = \ell_3$. 

As a consequence, 
the common center of mass for the equilateral solution 
remains unchanged. 
Without this unexpected thing, 
our calculations would become much more lengthy. 
The above expressions for the inner products 
are substituted into the R.H.S. of Eq. (\ref{EIH-EOM}). 
After straightforward calculations,  
the equation of motion for $M_1$ 
can be written as 
\begin{eqnarray}
- \omega^2 \boldsymbol{x}_1 
&=& - \frac{M}{a^3} \boldsymbol{x}_1 + g_{PN 1} \boldsymbol{x}_1 
\nonumber\\
&& 
+ 
\frac{\sqrt{3}M}{16 a^3} 
\boldsymbol{n}_{\perp 1} 
\frac{M_2M_3 (M_2 - M_3)}{M_2^2 + M_2M_3 + M_3^2}
\nonumber\\
&& 
~~~~~~~~~~~
\times
\bigg[10 + \frac{a^3}{M^2}\big(-4M_1 + 5M_2 + 5M_3\big) 
\omega^2 \bigg] , 
\label{M1-EOM}
\end{eqnarray}
where we used Eq. (\ref{omega2}) for velocity-dependent terms, 
$\boldsymbol{n}_{\perp 1} = \boldsymbol{v}_1/\ell_1\omega$ 
is defined as the unit normal vector to $\boldsymbol{x}_1$,  
and $g_{PN 1}$ denotes 
the post-Newtonian terms defined as 
\begin{eqnarray}
g_{PN 1} &=& 
\frac{1}{16 a^4 (M_2^2 + M_2M_3 + M_3^2)}
\nonumber \\
&&
\times
\Bigl[
80 (M_2^2 + M_2M_3 + M_3^2) M^2
-2 (8M_2^3 + M_2^2M_3 + M_2M_3^2 + 8M_3^3)M
 \nonumber\\
&& \mspace{30mu} 
- \Bigl\{ 
32 (M_2^2 + M_2M_3 + M_3^2) M^2 + 12M_2M_3 (M_2 + M_3) M
\nonumber \\
&& \mspace{50mu} 
- (16M_2^4 + 41M_2^3M_3 + 84M_2^2M_3^2 + 41M_2M_3^3 + 16M_3^4)
\Bigr\}
\frac{a^3}{M} \omega_N^2 \Bigr] .
\end{eqnarray}
Here, terms with $\omega_N^2$ come from 
the velocity-dependent terms and may be reexpressed 
by using Eq. (\ref{omega2}). 

We should note that the third term in the R.H.S. 
of Eq. (\ref{M1-EOM}) is parallel to 
the velocity of $M_1$ and thus perpendicular 
to $\mbox{\boldmath $x$}_1$ for a circular motion case. 
Therefore, the mass $M_1$ can be in circular motion, 
if and only if the coefficient of the third term vanishes, 
that is $M_2 = M_3$. 
Likewise, the masses $M_2$ and $M_3$ can be in circular motion, 
if and only if $M_3 = M_1$ and $M_1 = M_2$, respectively. 
Hence, all the three masses can have a circular motion, 
if and only if $M_1 = M_2 = M_3$. 

The previous paragraph postulates that three masses are finite. 
Here, a test mass is included. 
First, we consider a case that two of the masses are finite 
and one is zero, for instance $M_1 \to 0$. 
The third term in the R.H.S. of Eq. (\ref{M1-EOM}) vanishes, 
if and only if $M_2 = M_3$. 
Clearly, the corresponding terms of the equation of motion 
for $M_2$ and $M_3$ vanish in the limit as $M_1 \to 0$. 
Hence, the equilateral configuration is possible 
for an equal-mass binary with a test mass. 
Next, we consider a case that one mass is finite and 
the other two are zero, for instance 
$M_2 \to 0$ and $M_3 \to 0$. 
When we put $O(\varepsilon) \equiv O(M_2) = O(M_3)$, 
the third term in the R.H.S. of Eq. (\ref{M1-EOM}) 
becomes $O(\varepsilon)$ and 
thus vanishes as $\varepsilon \to 0$. 
The corresponding terms for $M_2$ and $M_3$ also 
become $O(\varepsilon)$ and 
thus vanish as $\varepsilon \to 0$. 
Therefore, the equilateral configuration is possible 
also for one finite mass and two test masses. 
This case is reasonable, since it is nothing but
two test masses orbiting around the Schwarzschild black hole 
(in the weak-field approximation). 

The remaining thing to do is to see whether orbital periods 
of the three masses are all the same 
in order to preserve the triangular shape 
if $M_1 = M_2 = M_3$. 
It is easy to see this, 
because one can obtain the post-Newtonian forces 
$g_{PN 2}$ and $g_{PN 3}$ from $g_{PN 1}$ by cyclic manipulations 
as $1 \to 2 \to 3 \to 1$, 
and finally by taking the equality of $M_1 = M_2 = M_3$, 
one can find $g_{PN 1} = g_{PN 2} = g_{PN 3}$. 
Therefore, it is concluded that 
the equilateral triangular configuration 
remains true for the post-Newtonian equation of motion 
in general relativity, 
if and only if all three masses are equal. 

Eq. (\ref{M1-EOM}) gives uniquely 
the post-Newtonian angular velocity as 
$\omega^2 = M a^{-3} - g_{PN}$, 
where $g_{PN} \equiv g_{PN 1} = g_{PN 2} = g_{PN 3}$ 
for $M_1 = M_2 = M_3 = M/3$. 
Here, $g_{PN}$ simply becomes 
\begin{equation}
g_{PN} = \frac{M}{a^3} 
\left( \frac{57}{12}\frac{M}{a} 
- \frac{41}{24} a^2 \omega_N^2 \right) . 
\label{gPN}
\end{equation}
One can show $g_{PN} > 0$ and hence $\omega < \omega_N$. 
This means that 
the angular velocity of the post-Newtonian equilateral triangular 
configuration is always smaller than the Newtonian one, 
provided that the masses and the side length are the same.
This behavior occurs also in the post-Newtonian collinear
configuration \cite{YA2011}.

\section{Summary}
We investigated the post-Newtonian effects on 
Lagrange's equilateral triangular solution 
for the three-body problem. 
For three finite masses, 
we found that the equilateral triangular configuration 
satisfies the post-Newtonian equation of motion 
in general relativity, 
if and only if all three masses are equal. 
When a test mass is included, 
the equilateral configuration is possible 
for two cases: 
(1) one mass is finite and the other two are zero, 
or 
(2) two of the masses are finite and equal, and 
the third one is zero, namely a symmetric binary with a test mass. 

It is left as a future work to examine post-Newtonian 
perturbations to triangular configurations for general masses. 
The configuration may be non-equilateral or 
non-periodic.

This work was supported in part (H.A.) 
by a Japanese Grant-in-Aid 
for Scientific Research from the Ministry of Education, 
No. 21540252.

\end{document}